\documentclass{aa}  

\usepackage{graphicx}
\usepackage{txfonts}
\usepackage{lipsum}
\usepackage{subcaption}         
\newcommand{\jwst}{\emph{JWST}\xspace}
\usepackage{lscape}             
\usepackage{placeins}           
                                
\usepackage[colorlinks=true,linkcolor=blue,citecolor=blue]{hyperref}
\newcommand{\ha}{H\(\alpha\)\xspace}
\newcommand{\hb}{H\(\beta\)\xspace}
\newcommand{\hg}{H\(\gamma\)\xspace}
\newcommand{\hd}{H\(\delta\)\xspace}
\newcommand{\oii}{[O\,{\sc ii}]\xspace}
\newcommand{\oiii}{[O\,{\sc iii}]\xspace}
\newcommand{\nii}{[N\,{\sc ii}]\xspace}
\newcommand{\neiii}{[Ne\,{\sc iii}]\xspace}

\newcommand{\nai}{\ion{Na}{i}\xspace}
\newcommand{\ki}{\ion{K}{i}\xspace}
\newcommand{\caii}{\ion{Ca}{ii}\xspace}
\newcommand{\rhat}{$\hat{\mathrm{R}}$\xspace}
\newcommand{\oiiidoublet}{[O\,{\sc iii}]\,4959,5007\,\AA\xspace}
\newcommand{\oxiii}{[O\,{\sc iii}]\,5007\,\AA\xspace}
\newcommand{\neoniii}{[Ne\,{\sc iii}]\,3869\,\AA\xspace}
\newcommand{\niired}{[N\,{\sc ii}]\,6583\,\AA\xspace}

\newcommand{\oiiiaur}{[O\,{\sc iii}]\,4363\,\AA\xspace}
\newcommand{\oiidoublet}{[O\,{\sc ii}]\,3726,3729\,\AA\xspace}

\newcommand{\msun}{M\(_\odot\)\xspace}

\begin{document}

\title{The metallicities of little red dot host galaxies:\\LRDs are metal poor, but not pristine}
\titlerunning{Little Red Dots are metal poor}

%
%
%

\author{G.~P.~Nikopoulos\inst{1,2}\fnmsep\thanks{georgios.nikopoulos@nbi.ku.dk}
        \and D.~Watson\inst{1,2}\fnmsep\thanks{darach@nbi.ku.dk}
        \and C.~L.~Pollock\inst{1,2}
        \and A.~Sneppen\inst{1,2}
        \and K.~E.~Heintz\inst{1,2,3}
        \and J.~Witstok\inst{1,2}
        \and G.~Brammer\inst{1,2}
        }

   \institute{Cosmic Dawn Center (DAWN), Denmark
            \and Niels Bohr Institute, University of Copenhagen, Jagtvej 155A, DK-2200, Copenhagen N, Denmark
            \and Department of Astronomy, University of Geneva, Chemin Pegasi 51, 1290 Versoix, Switzerland}

   \date{Received July XX, 2026}

 
  \abstract
    {Little Red Dots (LRDs) are a population of high-$z$ sources discovered by \jwst whose compactness, broad permitted lines, strong absorption features, continuum shapes and luminosities point to accreting supermassive black holes (SMBHs) embedded in dense gas. 
    To date, the metallicity of the hosts of these systems has not been systematically measured.}
    {We determine the gas-phase metallicities of LRD host galaxies and test whether their narrow-line emission is consistent with chemically pristine gas, metal poor star formation or AGN activity.}
    {We assemble a sample of 24 LRDs at $z \approx 2.3{-}7$ with  medium and high-resolution \jwst/NIRSpec data. We derive oxygen abundances and electron temperatures using the direct $T_\mathrm{e}$ method applied exclusively to the narrow components of emission lines, and cross-check against widely used strong line calibrations.}
    {We derive a sample-averaged abundance of $Z_{T_\mathrm{e}} = 0.08_{-0.03}^{+0.11}\,\mathrm{Z_{\odot}}$ ($T_\mathrm{e} = 23000_{-7000}^{+17000}$\,K), placing LRDs firmly in the metal-poor regime of high redshift star forming galaxies. The \rhat calibration yields a consistent average of $Z_{\hat{\mathrm{R}}} = 0.07_{-0.04}^{+0.07}\,\mathrm{Z_{\odot}}$, with only 4\% scatter relative to the direct $T_\mathrm{e}$ method, providing a robust proxy for systems where the \oiiiaur line is not detected. Using the strong-line method, we identify two extremely metal-poor LRDs with metallicities <1.3\%. }
    {The general population of LRDs are among the lower metallicity galaxies found by \jwst at this epoch and are an order of magnitude below metallicities of typical Seyfert or quasar host galaxies. They exhibit a narrow range of metallicities with a range of about 0.6\,dex, which remains remarkably stable over cosmic time. Such low metallicity may then be a defining property of the class. The fact that LRDs have substantial metallicity across most of the class poses a challenge to models that require formation via pristine gas collapse, while their generally low metallicity indicates that they are not standard AGN.}
    

   \maketitle
\nolinenumbers

\section{Introduction}
Little Red Dots are a new phenomenon discovered by \jwst \citep{Harikane2023,Matthee2024,Kocevski2024,Killi2024,Greene2024} mostly at \(z\sim4{-}7\), with a rapid drop-off at lower redshift \citep{Park2026,Lin2026-DESI}. They are spatially compact in the redder filters, have broad H and He lines \citep{Matthee2024,Greene2024,Juodzbalis2025} with exponential shapes \citep{Rusakov2025}, and a spectral break occurring close to the Balmer limit wavelength \citep{Setton2024,Akins2024cosmoswebb}. These properties have been variously argued to be consistent with tidal disruption events \citep{Bellovary2025}, late-stage quasi-stars \citep{Begelman2025}, or even supermassive stars \citep{Nandal2025}. In general, however, most models now involve accretion onto a supermassive black hole in a high density environment \citep{Inayoshi2025, Maiolino2024Xray}. This high density gas cocoon reprocesses radiation from the SMBH accretion, producing strong Balmer \citep{deGraaff2025b,Naidu2025,Sneppen2026_sirocco} and Paschen \citep{Sneppen2026b} features, as well as the electron-scattering broadened permitted lines commonly observed in LRDs \citep[see e.g.][]{Rusakov2025,Kokorev2025, Chang2025,Brazzini2026,Matthee2026}. This envelope of dense gas is also likely to suppress radio emission and X-rays \citep{Maiolino2024Xray,Sneppen2026c, Comastri2026, Tortosa2026}, while the high densities can also account for the anomalous Balmer emission observed in LRDs \citep{D'Eugenio2025deviation,Yan2025, Nikopoulos2026}. Radiative transfer simulations in multi-dimensional, moving gas have successfully reproduced most spectral features, indicating the need for a higher density, slow-moving inflow, and a faster outflow, with column densities of \(10^{24.5}{-}\,10^{25.5}\)\,cm\(^{-2}\) \citep{Sneppen2026_sirocco}. 

There is an ongoing debate on the precise nature of these sources, with interpretations including a stellar-like photosphere \citep[e.g.][]{Kido2025,Inayoshi2025b,deGraaff2025,deGraaff2025b}, reminiscent of a quasi-star interpretation \citep{Begelman2025}, a supermassive star \citep{Nandal2025} and perhaps related to direct-collapse black hole formation \citep{Pacucci2026}. However, evidence for both inflowing and outflowing gas in many objects \citep{Davis2026,Matthee2026} militates against a simple hydrostatic configuration. Moreover, strong emission lines emerging from within the scattering region are difficult to reconcile with a fully reprocessing photosphere. Conversely, producing the broad emission lines outside a cool reprocessed atmosphere with $T\!\lesssim\!6000$\,K is also challenging \citep{Martins2026}, while the observed line-to-continuum ratios are large and suggestive of a nebular-dominated continuum, even without an additional thermalised component \citep{Sneppen2026b}.

Recent studies of individual LRDs place them in a metal-poor regime \citep[$\sim\,$5--15\% $\mathrm{Z_{\odot}}$][]{Greene2024, Juodzbalis2024_rosetta,Killi2024, Tripodi2024, D'Eugenio2025deviation,Torralba2025,DEugenio2026,Ivey2026}, although \citet{Maiolino2025} argue for a very metal-poor counterpart in Abell2744-QSO1. 
Several observations further indicate that LRDs are not pristine systems, but are surrounded by high--column-density gas that is likely enriched: deep limits on soft X-rays require at least modest photoelectric absorption by metals \citep{Sneppen2026c}, while strong metal absorption features, including \caii, \nai, and \ki, point to substantial cool metal-enriched gas along the line of sight \citep{Lin2025,DEugenio2025b,Ji2025b}. The possible detection of water-vapour signatures in several low-redshift LRDs \citep{wang2026} suggests the presence of dense molecular gas with significant columns of oxygen surrounding these sources. 

Despite this mounting evidence for non-pristine, high column density gas, the metal abundances of LRDs and their hosts have not, to date, been systematically measured. The assumptions on the metallicity of LRDs greatly influence the interpretation of their spectra, as well as the results of SED and radiative transfer modelling. In addition, metallicity is a key discriminant between the proposed LRD formation channels, with pristine environments favouring a recent direct-collapse black hole formation scenario \citep{Pacucci2026}, whereas a measurement of substantial enrichment would instead point to a host that has already undergone significant star formation. Therefore, a robust abundance determination would put meaningful constraints on both future modelling of the observed conditions of LRDs and their evolution across cosmic time. 

In this work, we present the first catalogue of gas-phase metallicities and temperatures for a sample of LRD hosts observed with medium and high resolution JWST/NIRSpec spectroscopy. We derive metallicities through the classical direct $T_\mathrm{e}$ method, and cross-check them against widely used strong line calibrations in the literature (\rhat, R23, O32, Ne3O2 and N2). We also provide the first comparison between the results derived using grating and low resolution NIRSpec PRISM spectroscopy. Throughout the paper we adopt a solar oxygen abundance of 12+log(O/H)~=~8.69 \citep{Asplund2009} when converting to Z/Z$_{\odot}$. The paper is organised as follows. In Section~\ref{sec: observations} we describe the observational sample and its selection; in Section~\ref{sec: analysis} we present our line-fitting procedure and the direct $T_\mathrm{e}$ and strong line calibration frameworks; in Section~\ref{sec: results} we report our results on the temperatures and metallicities of LRDs, before discussing their implications on the nature of LRDs in Section~\ref{sec: discussion}. We summarise our findings in Section~\ref{sec: conclusion}.

\section{Observational samples}
\label{sec: observations}
We use data from the DAWN \jwst\ Archive (DJA) \citep{DeGraaff2024_RUBIES,Pollock2026,Heintz2025}, where  objects from previous LRD samples are included \citep[e.g.][]{Barro2025,deGraaff2025b,PerezGonzalez2026,Sneppen2026_sirocco,Matthee2026}. The data are publicly available from several \jwst observing programmes using the NIRSpec spectrograph \citep{Jakobsen2022} with the following Programme IDs: 1180, 1181, 1286 (JADES; \citealt{Eisenstein2023}), 1212 (GTO WIDE; \citealt{Maseda2024}) 9223 (PI: S. Fujimoto, \citealt{Atek2025}), 3567 (DEEPDIVE; \citealt{Ito2025}), 4233 (RUBIES; \citealt{deGraaff2025RUBIES}), 1345 (CEERS; \citealt{Finkelstein2023}), 4106 (PI: E. Nelson, \citealt{Nelson2023}), 4287 (PI: C. Mason, \citealt{Mason_survey,Whitler2026-4287}, LRD observations presented in \citealt{Tang2025-LRD4287}) and 9214 (SPURS; PIs: C. Mason \& D. Stark, \citealt{Chen2026-Spurs}).
We use the LRD sample to be presented in Nikopoulos et~al. (in prep.). This sample is the same as that used in the updated paper by \cite{Sneppen2026_sirocco}. 
In brief, this sample of LRDs is selected using color, compactness, and the presence of broad H$\alpha$ lines. 
The color selection requires either a significant Balmer break ($D_{4000}=F_{4200Å}/F_{3500Å}>1.2$, following \citealp{Binggeli2019}), which measures the strength of the hydrogen-related spectral discontinuity at 4000\,\AA, or a strong change in spectral slope between the ultraviolet and optical ($\beta_{\rm UV}<-0.2$, $\beta_{\rm opt}>0$, $\beta_{\rm opt}-\beta_{\rm UV}>0.5$ following \citealp{deGraaff2025b}). 
Compactness is assessed using S\'ersic profile fits in the \(F444W\) filter. The presence of broad emission lines is quantified using the Bayesian Information Criterion (BIC), requiring $\Delta \mathrm{BIC} > 10$ in favor of a model that includes a broad-line component.

\section{Methods}
\label{sec: analysis}

We interpret the Balmer emission in LRDs as consisting of an intrinsic line core plus exponential wings produced by electron scattering through a Compton-thick medium \citep{Rusakov2025}.  In this picture, the source is a central engine, such as an accreting supermassive black hole, embedded within a cocoon of highly dense, ionised gas \citep{Rusakov2025,Naidu2025,deGraaff2025Cliff,Inayoshi2025, Sneppen2026_sirocco}. 
We decompose the Balmer line profiles into a narrow component representing emission from the host galaxy (and/or a possible Narrow Line Region, NLR), and a broad component. The broad component consists of the sum of an `unscattered' and a `scattered' component, modelled respectively as a simple Gaussian, and a Gaussian with the same width convolved with an exponential. Where necessary, an absorption component is mutiplied by this model to account for the absorption lines frequently detected in LRD spectra.
For a more detailed analysis of the fitting model adopted, the reader is referred to \citet{Rusakov2025, Nikopoulos2026}. 

We use the fiducial model outlined above to fit the all the \ha lines in our sample. The width and redshift of the narrow Balmer components are tied to those of the [O\,{\sc iii}]\,4959,5007\AA\ doublet in velocity space, with each parameter having flat priors whose limits are dictated by the [O\,{\sc iii}]\,4959,5007\AA\ fit uncertainties. The broad \hb, \hg, and \hd profiles are scaled to match the respective components of the \ha profiles, except in rubies-egs61\_4233\_55604, rubies-egs63\_4233\_49140, and jades-gdn2\_1181\_28074, where the \hb broad components statistically prefer an independent fit.  All line widths are corrected for the instrumental resolution, for which we adopt an effective spectral resolution that is 1.7 times better than the nominal value reported in the NIRSpec documentation, following instrument modelling of point sources presented in the literature \citep{deGraaff2024}. Using a different effective resolution does not alter the fluxes inferred, and therefore the results of this work are largely unaffected by this choice, as shown in \citet{Nikopoulos2026}.

In this work, we use \texttt{dynesty} v.2.1.5 \citep{Speagle2020,sergey_koposov_2024_12537467} with static nested sampling \citep{Skilling2004, Skilling2006}, adopting $150n$ live points, where $n$ is the number of free parameters. To explore complex posterior distributions efficiently, we employ random-walk sampling \citep{Skilling2004} together with multi-ellipsoidal bounding \citep{Feroz2009} around the live points, which helps to capture degeneracies and multimodal structure. We then use the posterior samples from \texttt{dynesty} to compute flux distributions for all components and emission lines. The best fit fluxes represent the median of the retrieved distribution, while the uncertainties given are to the 16th and 84th percentiles. We consider a line/component to be detected when its flux exceeds zero by more than 3$\sigma$, while a 2$\sigma$ limit is quoted in case of a non-detection. We present the fluxes estimated in this work for all objects in Table~\ref{tab:fluxes}. All measurements presented are made by propagating each line flux's posterior probability distribution, using only narrow components/lines.  We did not attempt to determine the metal abundances using the broad lines, as the regions from which this emission arises are almost certainly above the critical densities of the forbidden metal lines.

\subsection{Direct $T_\mathrm{e}$: Deriving electron temperatures and elemental abundances}

The gas-phase metallicities were derived with a classical direct $T_\mathrm{e}$ framework using \texttt{PyNeb} \citep{pyneb1, pyneb2}. For each source, posterior probability flux distributions were assembled for the \oiiidoublet, \oiiiaur, \oiidoublet, and \hb narrow component. Electron temperatures were computed from the nebular-to-auroral oxygen ratio, $R_{\mathrm{OIII}}=\frac{I(4959)+I(5007)}{I(4363)}$, adopting fixed-density runs at a range of electron densities of $n_\mathrm{e}$ = $10^2$ - $10^6$ $\mathrm{cm^{-3}}$. All results quoted in this paper correspond to the case of $n_\mathrm{e} = 10^2\,\,\mathrm{cm^{-3}}$. The resulting $T_\mathrm{e}$ posteriors were retained per object, including an upper limit flag when the \oiiiaur line was not significantly detected. The posteriors were then propagated to compute oxygen abundances as $\frac{\mathrm{O}}{\mathrm{H}}=\frac{\mathrm{O}^{+}}{\mathrm{H}}+\frac{\mathrm{O}^{++}}{\mathrm{H}}$. 


For object IDs glimpse-obs01\_9223\_12248, glimpse-obs01b\_9223\_5536, rubies-uds3\_4233\_47509, there is no grating coverage of the \oiidoublet lines. These objects show temperatures greater than 20000\,K, however (see Table~\ref{tab:metallicity_te}). The average \oiiidoublet/\oiidoublet ratios of other similarly hot objects in the sample, for which the \oiidoublet doublet is detected, are $17^{+21}_{-10}$. Thus, for these objects, the O$^+$ ionization state was neglected, since its average contribution to the total oxygen abundance is expected to be minimal.

We note that we use all narrow Balmer line components available per object to estimate the narrow line $A_V$ using the \texttt{dust\_extinction} module \citep{Gordon2024-dust_ext} and assuming an SMC extinction curve \citep{Gordon2024-SMC}. We find that for most (though not all) objects, $A_V\!=\!0$ within $1\sigma$. (We caution, however, that narrow line Balmer decrements are poorly constrained due to degeneracies in the line-core decomposition of Balmer lines.) Other works also indicate minimal dust in the host galaxy emission \citep[e.g.][]{Killi2024, Matthee2024, Brooks2025, Zhang2025-NarrowLRDS, Nikopoulos2026}. We therefore only extinction-correct objects whose estimated narrow-component $A_V$ exceeds 0 at $> 3\sigma$. Specifically, we note that objects jades-gds06\_1286\_159717 and jades-gdn2\_1181\_28074 show $A_V = 1.4\pm 0.2$, and $A_V = 2.0 \pm 0.1$, respectively. A bigger systematic analysis of the dust emission of the full sample is left for future work.


\subsection{Strong line diagnostics}
\label{sec-analysis-strong line diagnostics}

We also used the estimated fluxes to calculate several strong line diagnostics, which are widely used in the literature to constrain the metallicities of high-$z$ star-forming galaxies, and are defined as follows:
\\

\begin{tabular}{@{}r@{\;}c@{\;}l} 
R23  &=& (\oiiidoublet + \oiidoublet)/ \hb, \\
O3 &=& \oxiii / \hb, \\
O2 &=& \oiidoublet / \hb, \\
O32 &=& \oxiii/ \oiidoublet, \\ 
Ne3O2 &=& \neoniii / \oiidoublet, and \\
$\hat{\mathrm{R}}$ &=&  0.47 log(O2) + 0.88 log(O3).\\  
\end{tabular}
\\

We compared our direct $T_\mathrm{e}$ metallicities against $\mathrm{R_{23}}$, the projected $\hat{\mathrm{R}}$ indicator, $\mathrm{O32}$, and $\mathrm{Ne3O2}$, using the relations from \citet{Scholte2025} and \citet{Sanders2026}. The $\hat{\mathrm{R}}$ and N2 = \niired/ \ha calibrations were also used to estimate the metallicity for all objects in the sample for comparison purposes, using the inverse relations from \citet{Scholte2025} and \citet{Sanders2026} respectively, while taking into account the intrinsic scatter of every relation. We used the $\hat{\mathrm{R}}$ calibration in \citet{Scholte2025}, as its low  metallicity (12 + log(O/H) < 7.2) extrapolation provides a more suitable fit to the estimated line ratios of similarly enriched objects. However, we note that the $\hat{\mathrm{R}}$ relation presented in \citet{Sanders2026} is completely consistent with the relation we used in this work for 6.8 < 12 + log(O/H) < 8.2, which is the regime that most LRDs in the sample lie in (see Sec. \ref{sec-results-temps and metallicities}).  We finally use N2 as an informative prior on the direct $T_\mathrm{e}$ metallicity, to truncate the high-O/H tail of the direct $T_\mathrm{e}$ posterior. 

\section{Results}
\label{sec: results}
\subsection{Direct $T_\mathrm{e}$ gas temperatures and metallicities}

\begin{figure}[h]
    \centering
    \includegraphics[width=0.49\textwidth]{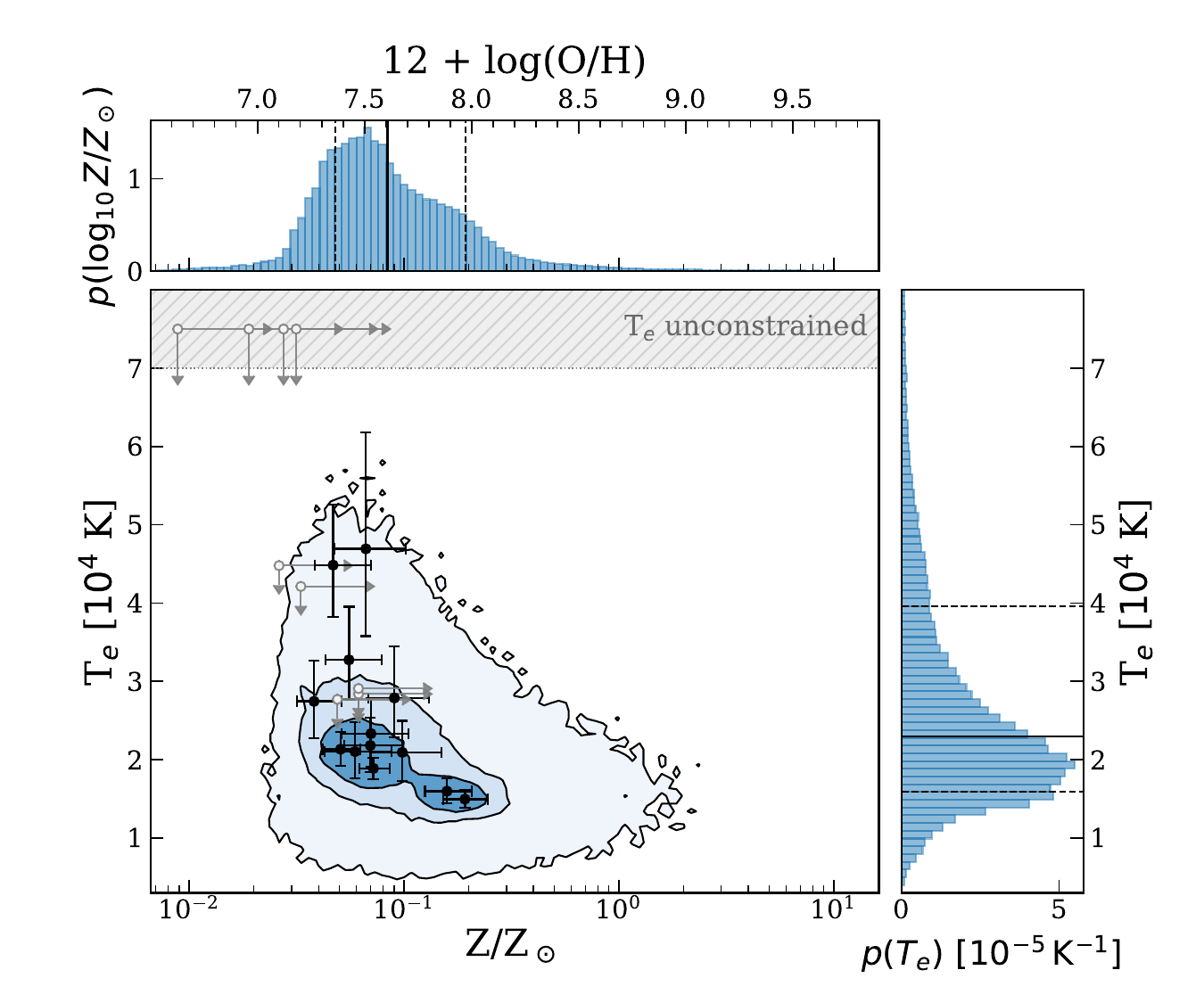}
       \caption{Metallicities and electron temperatures from the direct $T_\mathrm{e}$ method, for the full sample. The 39th, 68th and 95th percentile contours of the combined posterior probability distributions for the sample are shown in the central panel, with histograms of the one dimensional distributions shown on their respective axes. Median values are indicated with a solid line and 1$\sigma$ intervals with  dashed lines.}
          \label{fig_results: contours}
\end{figure}
Across the grating sample, we recover a broad range of electron temperatures $\mathrm{T}_\mathrm{e}\!\sim\!15\,000{-}45\,000$\,K. Specifically, we find that the modes of the derived temperature distributions for 18/22 direct $T_\mathrm{e}$ objects lie well below 30\,000\,K, in broad similarity with ISM conditions of high-redshift star-forming systems \citep[see e.g.][]{Laseter2024, Morishita2024, Chakraborty2025, Nishigaki2025, Pollock2026, Sanders2024,Sanders2026}. We note that a significant fraction of the sample lacks a robust \oiiiaur line detection; for those objects we quote 2$\sigma$ upper limits on the electron temperature, which translate into lower limits on the derived oxygen abundance.

\begin{figure}[b]
    \centering
    \includegraphics[width=0.48\textwidth]{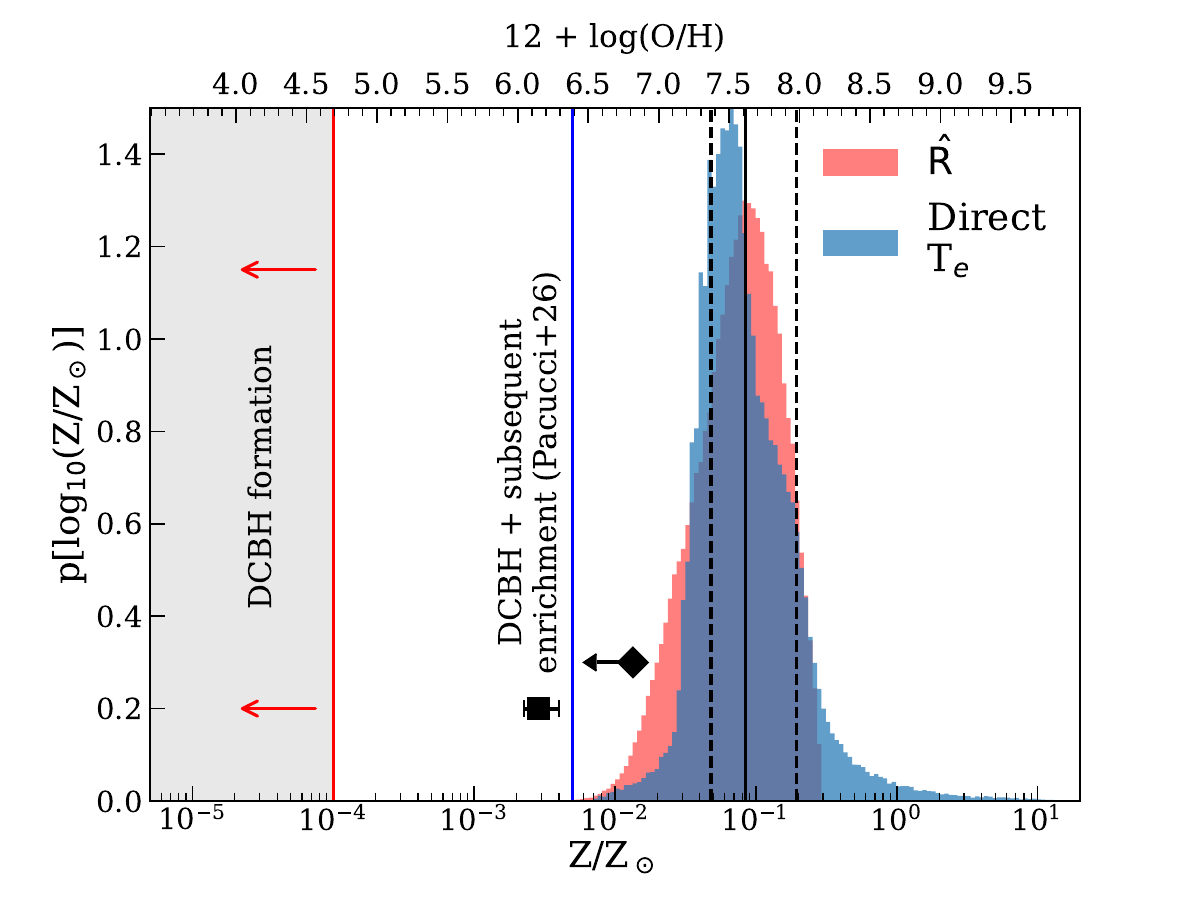}
       \caption{Summed posterior probability distributions of the metallicities determined using the direct $T_\mathrm{e}$ (blue) and ${\rm\hat{R}}$ (red) methods. Both methods suggest sub-solar metallicities with, $Z_{T_\mathrm{e}}=0.08_{-0.03}^{+0.11}\,\mathrm{Z}_\odot$ and $Z_{\rm\hat{R}}=0.07_{-0.04}^{+0.07}\,\mathrm{Z}_\odot$. \rhat metallicity estimates for abell2744-spurs02\_9214\_41 (diamond) and jades-gds-w09\_1212\_2993 (square) are shown separately and are placed in arbitrary y axis values. The theoretical limits for DCBH formation from near-pristine gas (red line) and DCBH formation plus subsequent enrichment to the LRD phase (blue line) are shown \citep{Pacucci2026}. Most of the sample is incompatible with these limits. }
          \label{fig_results: metallicity_posteriors}
\end{figure}

The resulting metallicities span a range of 0.05--0.2\,Z$_{\odot}$, indicating that gas in LRDs is metal-poor. We find that the mode of the metallicity distribution of the bottom 80\% the objects lies below 0.1\,Z$_{\odot}$, while the mode of the remaining 20\% is less than 0.2\,Z$_{\odot}$. The $T_\mathrm{e}$, as well as the metallicities derived for the full sample can be found in Table~\ref{tab:metallicity_te}.  We also plot a contour map of the combined derived $T_\mathrm{e}$ and metallicity distributions of all objects in Figure~\ref{fig_results: contours}, and we report an average $\mathrm{Z_{T_e}} = 0.08_{-0.03}^{+0.11}\,\mathrm{Z_{\odot}}$ and $T_\mathrm{e} = 23000_{-7000}^{+17000}$\,K for our sample. In Figure~\ref{fig_results: metallicity_posteriors}, we present the stacked direct $T_{\mathrm{e}}$ metallicity posteriors of the full sample. Our results then indicate that the gas in LRDs is indeed metal poor, but nowhere near pristine conditions. We have assumed a moderate density (\(100\,\mathrm{cm}^{-3}\)) throughout our analysis, in accordance with the range of densities inferred for LRDs at low $z$ \citep{Lin2026-DESI}. We note that since the oxygen abundance is dominated by the \oiiidoublet flux, which has a critical density of $n_\mathrm{e} \sim 7\times10^{5}\,\mathrm{cm^{-3}}$ \citep{Baskin2005}, increasing the density up to \(10^4\,\mathrm{cm}^{-3}\) does not alter our results. Different densities could not make the metal abundances significantly lower than we derive, so that our conclusions on non-pristine gas are robust. However the inferred abundance could increase by \(\sim0.3\)\,dex (a factor of two in solar units) if very high densities (\(10^5\,\mathrm{cm}^{-3}\)) are assumed for the narrow lines. While this seems extreme for the hosts of LRDs, there is some evidence of such high densities in three \oiii-emitting regions in star-forming galaxies at \(z=6.1\), 6.1, and 7.0,  \citep{Arellano-Cordova2026}, so such high densities cannot be ruled out at present. On the other hand, even allowing for a 0.3\,dex upward shift in the \(z\gtrsim5\) objects in our sample would not greatly affect our conclusions. Furthermore, we detect significant \oiidoublet emission in nine objects out of 22 measured in the sample. At \(10^5\,\mathrm{cm}^{-3}\) the \oiidoublet would be strongly suppressed. The measurements we have, give O32 ratios and lower bounds between 4 and 20, which suggest a density for a single-zone medium that is not far above the critical density for \oiidoublet, \(\sim10^4\,\mathrm{cm}^{-3}\). Of course a multi-zone medium would allow this constraint to be evaded.

\begin{table*}[t]
\centering
\renewcommand{\arraystretch}{1.3}
\setlength{\tabcolsep}{5pt}
\caption{Estimated temperatures and metallicities for all methods used.}\label{tab:metallicity_te}
\resizebox{1\textwidth}{!}{%
\begin{tabular}{@{}clccccccc@{}}
\hline
ID & Object & $z$ & $12+\log(\mathrm{O/H})$ [T$_\mathrm{e}$] & $\mathrm{T_e }$ [K] & $12+\log(\mathrm{O/H})$ [$\hat{\mathrm{R}}$]  &$12+\log(\mathrm{O/H})$ [N2] & DT$_\mathrm{e}$+N2 &Z/$\mathrm{Z_{\odot}}$ \\
\hline
1 & abell2744-spurs02\_9214\_41 & $7.037$ & --- & --- & $<6.81$ & --- & --- & --- \\
2 & rubies-egs61\_4233\_55604 & $6.983$ & $7.36_{-0.08}^{+0.17}$ & $44800_{-6600}^{+7700}$ & $7.46_{-0.25}^{+0.29}$ & --- & $7.36_{-0.08}^{+0.17}$ & $0.05_{-0.01}^{+0.02}$ \\
3 & rubies-uds42\_4233\_807469 & $6.775$ & $>7.13$ & $<74000$ & $7.29_{-0.26}^{+0.30}$ & --- & $> 7.13$ & $> 0.03$ \\
4 & jades-gdn2\_1181\_954 & $6.760$ & $7.54_{-0.14}^{+0.18}$ & $23300_{-4300}^{+4600}$ & $7.41_{-0.23}^{+0.26}$ & $<8.10$ & $7.55_{-0.14}^{+0.17}$ & $0.07_{-0.02}^{+0.04}$ \\
5 & egs-nelsonx\_4106\_47962 & $6.728$ & $7.46_{-0.14}^{+0.17}$ & $21000_{-3500}^{+3800}$ & $7.28_{-0.21}^{+0.23}$ & $<8.18$ & $7.48_{-0.15}^{+0.17}$ & $0.06_{-0.02}^{+0.03}$ \\
6 & rubies-egs63\_4233\_49140 & $6.685$ & $7.51_{-0.15}^{+0.19}$ & $46900_{-11200}^{+14900}$ & $7.59_{-0.25}^{+0.26}$ & $<7.98$ & $7.50_{-0.14}^{+0.17}$ & $0.06_{-0.02}^{+0.03}$ \\
7 & glimpse-obs01b\_9223\_5536 & $6.223$ & $>7.19$ & --- & $7.71_{-0.22}^{+0.22}$ & --- & $>7.19$ & $>0.03$ \\
8 & glimpse-obs01\_9223\_12248 & $6.108$ & $7.39_{-0.08}^{+0.09}$ & $21300_{-2100}^{+2200}$ & $7.63_{-0.19}^{+0.22}$ & $<8.06$ & $7.40_{-0.09}^{+0.09}$ & $0.05_{-0.01}^{+0.01}$ \\
9 & jades-gdn\_1181\_38147 & $5.869$ & $7.53_{-0.12}^{+0.15}$ & $21800_{-3400}^{+3500}$ & $7.61_{-0.20}^{+0.22}$ & $<8.23$ & $7.69_{-0.19}^{+0.20}$ & $0.10_{-0.03}^{+0.06}$ \\
10 & rubies-uds3\_4233\_47509 & $5.672$ & $>7.38$ & $<27700$ & $7.78_{-0.19}^{+0.19}$ & $<7.80$ & $7.71_{-0.21}^{+0.22}$ & $0.10_{-0.04}^{+0.07}$ \\
11 & ceers\_1345\_746 & $5.622$ & $>6.64$ & --- & $>7.21$ & $<8.46$ & $7.22_{-0.36}^{+0.53}$ & $0.03_{-0.02}^{+0.08}$ \\
12 & jades-gds05\_1286\_204851 & $5.482$ & $>7.48$ & $<29100$ & $7.68_{-0.23}^{+0.22}$ & $<7.98$ & $7.82_{-0.21}^{+0.22}$ & $0.14_{-0.05}^{+0.09}$ \\
13 & ceers-egs\_1345\_2782 & $5.239$ & $>7.21$ & $<42200$ & $7.33_{-0.29}^{+0.33}$ & --- & $>7.21$ & $>0.03$ \\
14 & jades-gds06\_1286\_159717 & $5.077$ & $7.68_{-0.15}^{+0.18}$ & $20900_{-3700}^{+4000}$ & $7.82_{-0.19}^{+0.17}$ & $<7.94$ & $7.68_{-0.14}^{+0.15}$ & $0.10_{-0.03}^{+0.04}$ \\
15 & jades-gdn\_1181\_68797 & $5.039$ & $7.43_{-0.11}^{+0.15}$ & $32700_{-5100}^{+6800}$ & $7.78_{-0.20}^{+0.19}$ & $<8.38$ & $7.73_{-0.27}^{+0.27}$ & $0.11_{-0.05}^{+0.09}$ \\
16 & valentino-egs\_3567\_42232 & $4.952$ & $>6.97$ & --- & $7.33_{-0.28}^{+0.32}$ & $<8.00$ & $7.31_{-0.21}^{+0.26}$ & $0.04_{-0.02}^{+0.03}$ \\
17 & jades-gdn\_1181\_39353 & $4.849$ & $>7.11$ & $<44900$ & $7.21_{-0.26}^{+0.30}$ & $<7.92$ & $7.49_{-0.26}^{+0.31}$ & $0.06_{-0.03}^{+0.07}$ \\
18 & jades-gds-w09\_1212\_2993 & $4.824$ & --- & --- & $6.15_{-0.10}^{+0.15}$ & $<8.24$ & --- & --- \\
19 & jades-gds02\_1286\_38562 & $4.821$ & $7.89_{-0.10}^{+0.11}$ & $16000_{-1600}^{+1600}$ & $7.67_{-0.20}^{+0.22}$ & $<7.90$ & $7.86_{-0.09}^{+0.10}$ & $0.15_{-0.03}^{+0.04}$ \\
20 & ceers-egs\_1345\_1244 & $4.477$ & $7.97_{-0.10}^{+0.11}$ & $14900_{-1100}^{+1200}$ & $7.67_{-0.21}^{+0.22}$ & $8.12_{-0.11}^{+0.09}$ & $8.05_{-0.08}^{+0.07}$ & $0.23_{-0.04}^{+0.04}$ \\
21 & jades-gdn\_1181\_73488 & $4.133$ & $7.27_{-0.08}^{+0.13}$ & $27400_{-4700}^{+5200}$ & $7.07_{-0.19}^{+0.19}$ & $<8.12$ & $7.32_{-0.11}^{+0.17}$ & $0.04_{-0.01}^{+0.02}$ \\
22 & jades-gds-wide\_1180\_13329 & $3.938$ & $>7.48$ & $<28400$ & $7.81_{-0.20}^{+0.18}$ & $<8.21$ & $7.88_{-0.20}^{+0.21}$ & $0.16_{-0.06}^{+0.10}$ \\
23 & jades-gdn\_1181\_53501 & $3.429$ & $7.65_{-0.12}^{+0.16}$ & $27900_{-5000}^{+6600}$ & $7.67_{-0.26}^{+0.24}$ & $<7.86$ & $7.63_{-0.12}^{+0.14}$ & $0.09_{-0.02}^{+0.03}$ \\
24 & jades-gdn2\_1181\_28074 & $2.260$ & $7.55_{-0.07}^{+0.07}$ & $18900_{-1400}^{+1300}$ & $7.63_{-0.19}^{+0.22}$ & $7.64_{-0.09}^{+0.06}$ & $7.59_{-0.06}^{+0.06}$ & $0.08_{-0.01}^{+0.01}$ \\
\hline
\end{tabular}%
}
\tablefoot{Abundances were estimated assuming an electron density of $10^2\,\mathrm{cm^{-3}}$. Upper limits on the temperature arise if the \oiiiaur line is not detected. Upper limits exceeding $10^5\,\mathrm{K}$ are marked with `---'. Objects with no \niired and \ha narrow component detections have no meaningful N2 estimate, and are also denoted with  `---'. The final column quotes the combination of the direct $T_\mathrm{e}$ and N2 metallicities in solar units.}
\end{table*}

We also find objects with unconstrained temperature upper limits, due to a combination of relatively weak \oiiidoublet emission, and a relatively high \oiiiaur line $2\sigma$ upper limit. These flux limits then yield temperature ranges that extend into the asymptotic regime of the \oiiidoublet / \oiiiaur vs $T_e$ relation \citep{OsterbrockFerland2006}. This is due to the limitations of the data with respect to the detection of the \oiiiaur line. For these objects, we use the \rhat calibration, as a proxy for their metallicity, since it provides a measurement that is in general consistent with the direct $T_\mathrm{e}$ method (see Fig.~\ref{fig_results: metallicity_posteriors}).

\label{sec-results-temps and metallicities}

\subsection{Strong line calibration metallicities}
\label{sec-results-strong line diagnostics}

\begin{figure*}
    \centering
    \includegraphics[width=\textwidth]{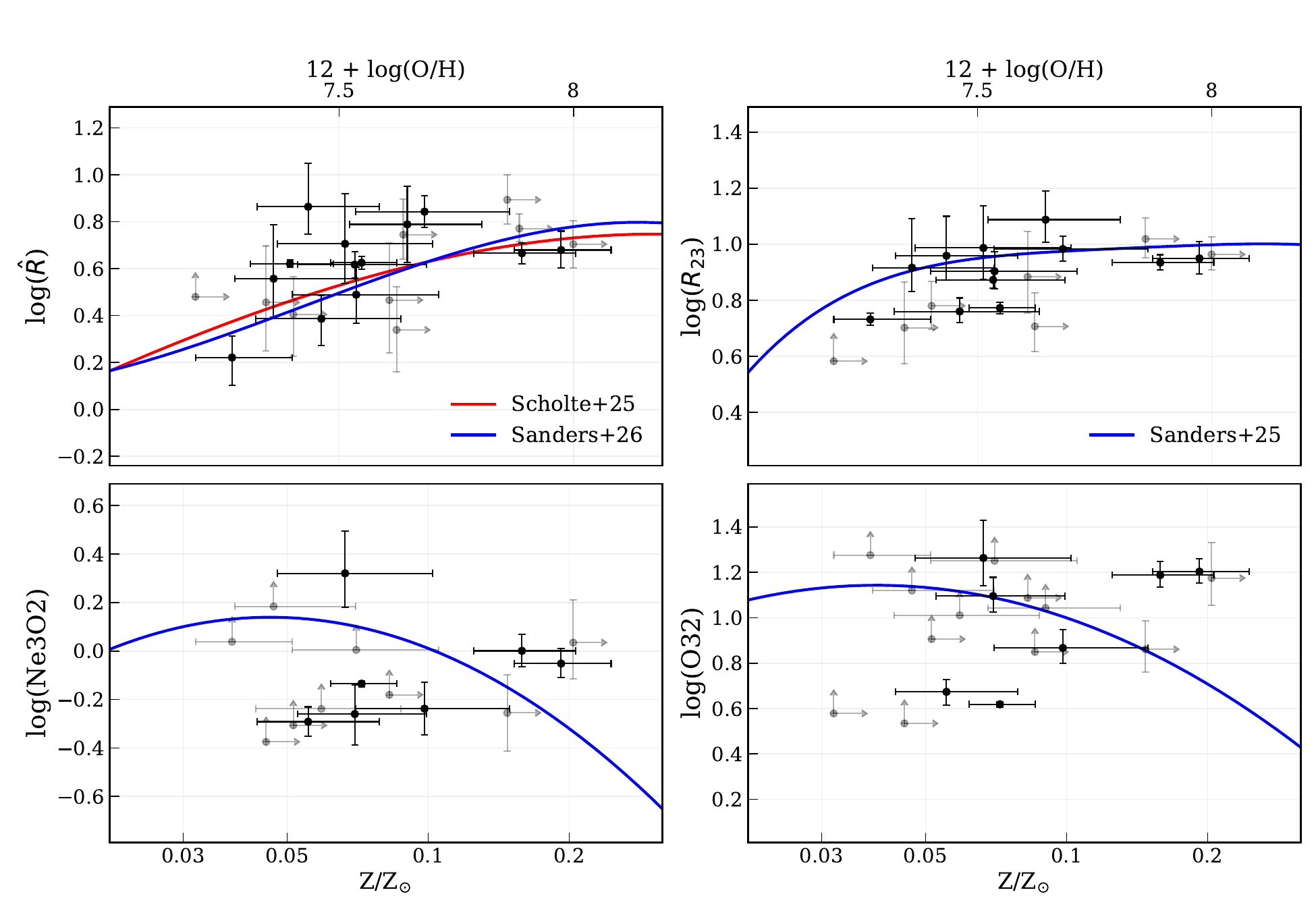}
       \caption{ Strong line calibration ratios vs.\ direct-$T_\mathrm{e}$ metallicities. We also plot the empirical calibrations from \citealt{Scholte2025} (red) and \citealt{Sanders2026} (blue) derived using high-$z$ star forming systems with \oiiiaur detections.}
          \label{fig_results: strong line calibrations}
\end{figure*}

Comparing our direct $T_\mathrm{e}$ metallicities against the widely used strong-line diagnostics shows that the bulk of our sample is broadly consistent with these calibrations (see Fig.~\ref{fig_results: strong line calibrations}). $\mathrm{R_{23}}$ and $\hat{\mathrm{R}}$ track the direct $T_\mathrm{e}$ values most tightly, reflecting their weaker sensitivity to the ionization state of the gas, whereas $\mathrm{O32}$ and $\mathrm{Ne3O2}$ display noticeably larger scatter.

The tightest and therefore most informative correlation is found with the ${\rm \hat{R}}$ diagnostic both for the population (Fig. \ref{fig_results: metallicity_posteriors}), as well as for individual objects (see top-left panel, Fig. \ref{fig_results: strong line calibrations}). The bulk of the population lies in the lower $\hat{\mathrm{R}}$-metallicity branch (\(Z < 0.3\,\mathrm{Z}_{\odot}\)) avoiding the ambiguity of the high-metallicity branch.  We thus derive the metallicities of LRDs using the $\hat{\mathrm{R}}$ calibration, and plot the combined  $\hat{\mathrm{R}}$ metallicity posteriors for all objects in the sample, in Fig.~\ref{fig_results: metallicity_posteriors}. We compute an average $Z_{\hat{\mathrm{R}}} = 0.07_{-0.04}^{+0.07}\,\mathrm{Z_{\odot}}$, in agreement with the average derived for the direct $T_\mathrm{e}$ method. It is noted that the highest likelihood $\hat{\mathrm{R}}$ metallicities are marginally higher than their direct $T_\mathrm{e}$ counterparts. As discussed in Section \ref{sec:discussion-sf-or-agn}, LRDs have temperatures that are consistent with the hotter branch of high $z$ SF galaxies. Since the $\hat{\mathrm{R}}$ vs. metallicity relation has been calibrated against such systems, a high redshift galaxy with the same $\hat{\mathrm{R}}$ value can be expected to have marginally lower temperature and therefore higher metallicity than a typical LRD. Nevertheless, the metallicities inferred are not affected, since 11/13 objects have consistent direct $T_\mathrm{e}$ and $\hat{\mathrm{R}}$ metallicities within $1\sigma$, while the remaining two are reconciled at $<2\sigma$. 
The insignificance of that effect is further corroborated by the fact that the $\hat{\mathrm{R}}$ metallicity posterior does not have a very high metallicity tail, as seen in the direct $T_\mathrm{e}$ posterior. This direct $T_\mathrm{e}$ tail is driven by non-detections of the \oiiiaur line, leading to low inferred temperatures, and thus high metallicities. The $\hat{\mathrm{R}}$ calibration is better constrained in that sense, since it does not depend on the usually faint \oiiiaur line. We therefore argue that the $\hat{\mathrm{R}}$ can be safely used to estimate the metallicity of LRDs, if the direct $T_\mathrm{e}$ method is not feasible.

The \niired line is not detected in 22 out of 24 objects in the sample; the N2-derived abundances are then quoted as upper limits where the \ha narrow component is detected, and are found to be broadly consistent with the direct $T_\mathrm{e}$ method abundances. N2 is strongly sensitive to the ionization parameter \citep{MoralesLuis2014} and to any AGN contribution to the flux of the \niired line, while star-forming galaxies at high redshift show moderate to extreme nitrogen enrichment \citep[e.g.][]{Bunker2023,Cameron2023,Isobe2023, Zhang2026-Nitrogen,Cataldi2025b, Cameron2026}. As a consequence, N2-based abundances are conservative when used as an upper bound, and can be used to independently constrain the high O/H tail of our direct $T_\mathrm{e}$ posteriors. Indeed, the combined direct $T_\mathrm{e}$ and N2 posteriors show on average lowered upper bounds, yielding $\mathrm{Z}_{\mathrm{joint}} = 0.08_{-0.04}^{+0.09}\,\,\mathrm{Z_{\odot}}$ for the full sample. 
The above population average is consistent with the values obtained using both the direct $T_\mathrm{e}$ method and the $\hat{\mathrm{R}}$ calibration. The results of all methods presented in this work can be found in Table~\ref{tab:metallicity_te}.

\subsection{Evidence for extremely metal poor LRDs}
\label{sec:results-pristine-LRDs}
In this section we highlight two objects in the sample, abell2744-spurs02\_9214\_41 (A2744-QSO1, $z = 7.04$; \citealt{Maiolino2025} and jades-gds-w09\_1212\_2993 ($z = 4.88$, \citealt{deGraaff2025b}). 

Abell2744-spurs02\_9214\_41 has been extensively discussed in the literature \citep{Maiolino2025, Juodzbalis2026-Qso1, Tang2026-qso1-spurs} and was reported to have \(Z = 0.0047\,\mathrm{Z_{\odot}}\) in its central region, while an extended region of 200\,pc around its core shows \(Z < 0.0039\,\mathrm{Z_{\odot}}\) \citep{Maiolino2025}. Indeed, we report that in this object all oxygen lines are non-detections, making a direct $T_\mathrm{e}$ measurement impossible. We then use the \rhat strong line calibration, and put a 2$\sigma$ upper limit on its metallicity of \(Z < 1.3\%\,\mathrm{Z_{\odot}}\), consistent with previous findings. 

Jades-gds-w09\_1212\_2993 shows no detected emission line at 3$\sigma$ apart from \ha, in all grating spectra available. We nevertheless use the \ha profile derived from fitting the available G395H spectrum in the DJA, to decompose the \hb+\oiiidoublet complex in the PRISM spectrum, which has much higher SNR. We are then able to estimate the narrow \hb and \oxiii fluxes, and thus use the O3 strong line calibration to estimate a metallicity of \(Z = 0.0034^{+0.0010}_{-0.0008}\,\mathrm{Z_{\odot}}\). This is consistent with the corresponding \rhat measurement, \(Z = 0.0029^{+0.0011}_{-0.0006}\,\mathrm{Z_{\odot}}\). This result firmly puts Jades-gds-w09\_1212\_2993 as one of the most metal poor LRDs discovered to date, with significant implications on the formation scenarios of LRDs (Fig.~\ref{fig_results: metallicity_posteriors}). We nevertheless caution that the use of PRISM low resolution spectra, under incorrect assumptions or certain redshift regimes, can lead to very different results when compared to the full direct $T_\mathrm{e}$ method, utilizing medium/high resolution spectra (see Sec.~\ref{sec:results-prism-comparison}).



\subsection{Comparison of grating- and PRISM-derived metallicities}
\label{sec:results-prism-comparison}

\begin{figure}
    \centering
    \includegraphics[width=0.45\textwidth]{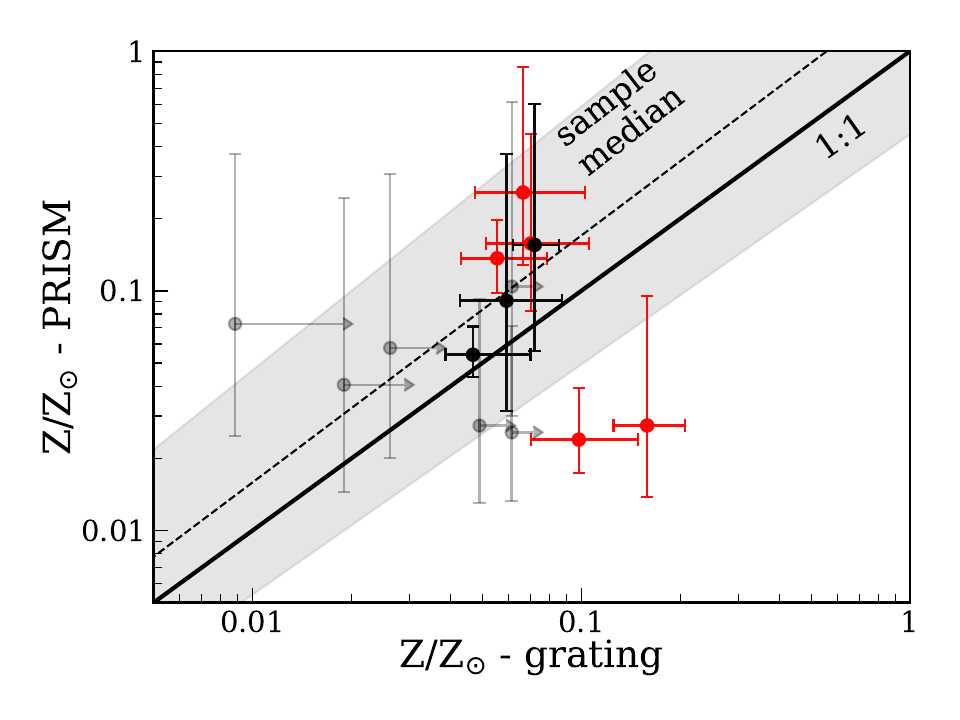}
       \caption{ Direct $T_\mathrm{e}$ metallicities from PRISM spectra compared to direct $T_\mathrm{e}$ metallicities from grating spectroscopy. The solid line shows the region on which the two metallicities are equal. The dashed line and shaded region correspond to the average observed PRISM/grating metallicity relation and 1$\sigma$ scatter of \(1.03\!\pm\!0.07\)\% in 12+log(O/H) space. Faded gray objects correspond to those whose grating metallicity estimation is a lower limit due to the non-detection of the \oiiiaur line. Data points coloured red signify objects for which the two measurements are not in agreement within $1\sigma$.}
          \label{fig_results: prism comparison}
\end{figure}

In this subsection, we test whether the results derived in this work match with abundances derived using NIRSpec PRISM spectra. To this end, we perform a very simple analysis of the PRISM spectra of objects in our sample which have medium/high-resolution-derived metallicities, in which 1) we fit Balmer lines with broad components which do not take into account any  quantities derived using grating spectra, 2) we neglect the possibility of a broadened \oiiiaur line, as found in rubies-egs63\_4233\_49140 \citep{Nikopoulos2026, D'Eugenio2025deviation} and 3) we assume all lines are detected. The latter criterion is forced so that the methodology emulates the results of automated fitting routines, like \texttt{msaexp} \citep{Brammer2023}. We then perform the same method detailed in Section~\ref{sec: analysis} to derive the PRISM-based metallicities, and compare them with the grating-based metallicities in Figure~\ref{fig_results: prism comparison}.

In the sample of 8 objects for which the metallicity reported is not a lower limit, and for which PRISM spectra are available, 5 out of 8 objects show an inconsistency between the values derived using grating and PRISM spectra. For 3 out of 5 objects, this inconsistency is reconciled at $2\sigma$, while jades-gdn\_1181\_68797 and jades-gds06\_1286\_159717 are reconciled at $3\sigma$. 


The above inconsistencies suggest that PRISM spectra might not have sufficient resolution to accurately decompose the \hb + \oiiidoublet, and the \hg + \oiiiaur complexes, and retrieve the various fluxes safely. In order to quantify the deviation between the two sets of measurements, we calculate the distribution of ratios of the oxygen abundance derived using PRISM, $12+\mathrm{log(O/H)}_{\mathrm{PRISM}}$ over the one derived using grating spectra, $12+\mathrm{log(O/H)}_{T_\mathrm{e}}$, for objects with \oiiiaur line detections in the grating spectra. We find that the median ratio of the distribution is indeed 1.03, with a $1\sigma$ scatter of 7\%. This effectively means that for an object with the average metallicity of the sample, 12 + log(O/H) = 7.59 (Z = 0.08 $\mathrm{Z_{\odot}}$), the PRISM estimate can be off by up to 0.8 dex at $1\sigma$, leading to possible measurements anywhere between 0.05\,Z$_{\odot}$ and 0.54\,Z$_{\odot}$. For comparison purposes, the median and scatter of the equivalent distribution of (12+log(O/H)$_\mathrm{\hat{R}}$)/(12+log(O/H)$_{T_\mathrm{e}}$) is $1\pm\,4\%$. We note that if one includes a $3\sigma$ non-detection threshold in the PRISM analysis, there are only 3 objects which have both constrained grating and PRISM metallicity measurements, with all 3 being inconsistent. The process of deriving the median ratio and scatter between the two sets, described above, is independent on whether a line is treated as an upper limit or not, and therefore holds irrespective of the detection threshold adopted.

We argue then that PRISM data could yield a sufficiently reliable metallicity estimate on individual objects, assuming that the resolution and SNR are sufficient to resolve the blends of Balmer and oxygen lines, and that the analysis is sophisticated enough and tailored to the object under investigation. On the other hand, caution should be exercised when using metallicity estimates of PRISM-based LRD sample studies, unless their analyses fulfill the above criteria.  

\section{Discussion}
\label{sec: discussion}

\subsection{On the metallicity of LRDs}

In this work, we provide the first metallicity estimate which uses medium and high resolution spectra for a sample of 24 LRDs. Our sample-averaged direct-$T_\mathrm{e}$ abundance of $\mathrm{Z = 0.08^{+0.11}_{-0.03}\,Z_\odot}$ (Sec.~\ref{sec-results-temps and metallicities}) places the LRDs firmly in the metal-poor, but not pristine, regime of high-redshift star-forming galaxies. The electron temperatures span $T_\mathrm{e} \approx 15000{-}45000$\,K, with the lowest values ($\sim 16000{-}21000$\,K) corresponding to the most enriched objects (jades-gds02\_1286\_38562; 12+log(O/H) = $7.89_{-0.10}^{+0.11}$ and ceers-egs\_1345\_2782; 12+log(O/H) = $7.97_{-0.10}^{+0.11}$), and the highest ($\gtrsim$ 30000K) generally reflecting more metal poor conditions (e.g. rubies-egs61\_4233\_55604 at $z$ = 6.98 with 12 + log(O/H) = $7.36^{+0.13}_{-0.08}$). The above metallicity regime overlaps with values measured using the direct $T_\mathrm{e}$ method, reported for individual high $z$ LRDs (e.g.\ \citealt{Juodzbalis2024_rosetta}; jades-gdn2\_1181\_28074 at $\sim 0.15\, \mathrm{Z_{\odot}}$ and \citealt{D'Eugenio2025deviation}; rubies-egs63\_4233\_49140 at roughly 8\% $\mathrm{Z_{\odot}}$ although with hints of a metallicity gradient due to density stratification). The derived population properties are also consistent with metallicity estimates for individual objects with grating spectroscopy that use strong-line calibrations. Specifically, \citet{Ivey2026} report 12+log(O/H) $\approx 6.93^{+0.07}_{
-0.09}$ for the Cliff at $z = 3.55$, and \citet{Maiolino_smallvigorous_2024} report 12 + log(O/H) $\approx 6.4$ for A2744-QSO1 at $z = 7.04$ 
, which lie at, and below the metal-poor tail of our distribution respectively. We finally report broad consistency with the metallicities derived for LRDs from other PRISM-based analyses in the literature \citep[e.g.][]{Killi2024, Greene2024, Tripodi2024}.

\begin{figure*}[h!]
    \centering
    \includegraphics[width=\textwidth]{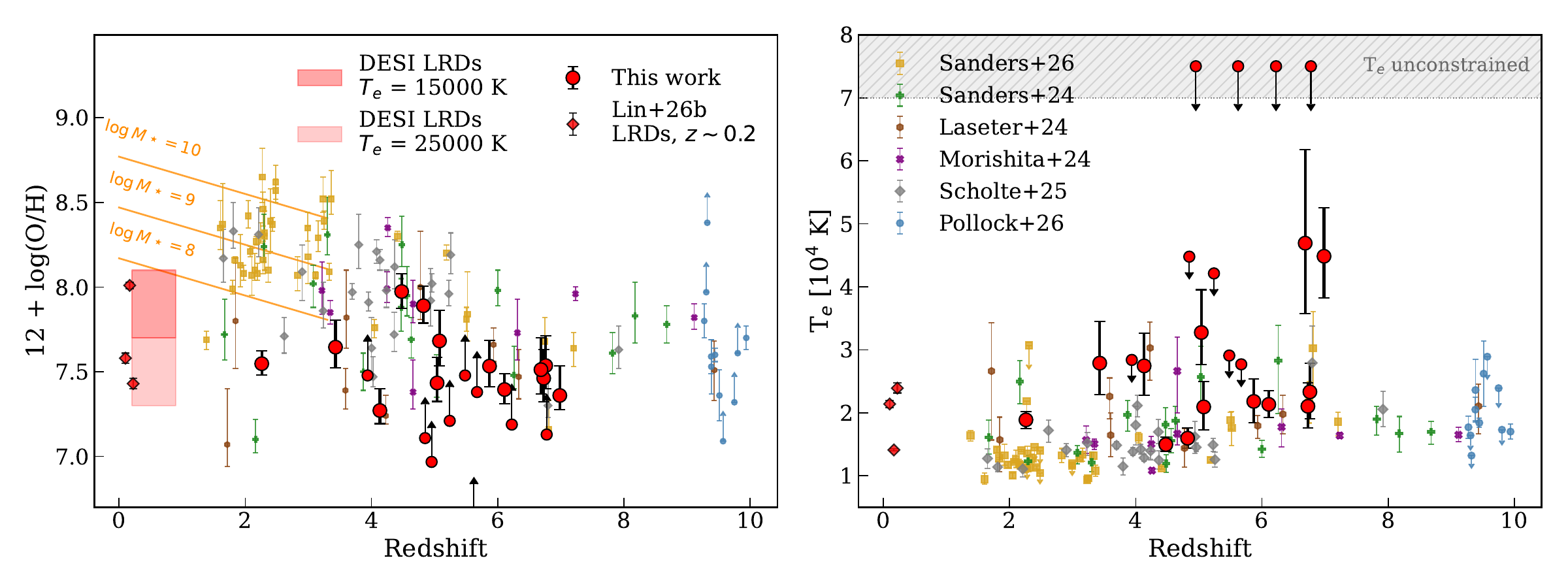}
       \caption{Metallicities (left) and electron temperatures (right) of our LRD hosts as a function of redshift. Metallicities are inferred from the direct $T_\mathrm{e}$ method. The three confirmed low-$z$ LRDs \citep{Lin2025} are also shown, as well as shaded regions corresponding to the metallicity distribution of the LRDs in the sample of \citet{Lin2026-DESI} from \(0.2<z<0.95\) assuming their assumed $T_\mathrm{e}\!=\!15\,000$\,K (dark pink) and a temperature most consistent with the LRDs in our sample, $T_\mathrm{e}\!=\!25\,000$\,K (light pink). The high-$z$ star-forming galaxies from \citet{Sanders2024,Sanders2026, Laseter2024, Morishita2024, Scholte2025,Pollock2026}, are also plotted for comparison. The orange solid lines correspond to the \citet{Sanders2021} MZR predictions for \(\log{M_*/\mathrm{M}_\odot}\!=\!8{-}10\), across the redshift range \(0<z<3\). The LRD hosts have low metallicities, even compared to hosts at similarly high redshift, and seem to have consistently low metallicity even at quite low redshifts.}
          \label{fig_discussion: redshift evolution}
\end{figure*}

By comparison, the metallicities of the low-redshift LRD analogues of \citet{Lin2025} are similar: three objects show metallicities and temperatures at $0.05{-}0.2\,\mathrm{Z}_{\odot}$ and $T_\mathrm{e} \sim 14100{-}24000\,$K, estimated using the direct $T_\mathrm{e}$ method. \citet{Lin2026-DESI} also report 27 LRDs at $z \sim 0.2{-}0.9$, and argue that their narrow line ratios indicate a mixture of star formation and AGN emission at that redshift ($T_\mathrm{e} > 20000$ K). Their measurements suggest metallicities in the range 10\%--25\%\,$\mathrm{Z_{\odot}}$, with a population average of 13\%\,$\mathrm{Z_{\odot}}$, assuming a constant $T_\mathrm{e} = 15000$\,K. We note that the results of this work (see Fig.~\ref{fig_results: contours}) suggest that the typical temperatures of LRD hosts may be higher; by adopting $T_\mathrm{e} \sim 25000$\,K instead, the metallicities of the \citet{Lin2026-DESI} DESI sample would drop to a range of 5\%--10\%\,$\mathrm{Z_{\odot}}$, as suggested by Fig.\ref{fig_results: contours}, bringing them in line with the bulk of the sample presented in this work. The consistency of the results of this work with LRDs at $z < 0.9$, suggests that there is minimal evolution in the enrichment of LRDs across cosmic time (see Fig.~\ref{fig_discussion: redshift evolution}). 

Several formation scenarios for LRDs have invoked initially pristine gas conditions, such as required for direct collapse black hole formation, $Z/\mathrm{Z}_{\odot}\lesssim10^{-5}{-}10^{-4}$ \citep{Omukai2008,Cenci2025}. Subsequent enrichment could increase the metallicity up to $5\times10^{-3}\,\mathrm{Z}_\odot$ \citep{Pacucci2026}. Importantly, only 2/24 objects show metallicities naively in agreement with such a scenario (although see caveats discussed in Sec.~\ref{sec:results-pristine-LRDs}). These objects represent just 8\% of the sample. Other results in the literature are in agreement with the population average found in this work for the direct $T_\mathrm{e}$ method, and seem to be inconsistent with the hypothesis that LRDs exist in extremely metal-poor or even pristine environments. However, these measurements only disfavor models in which the actively observed LRD phase occurs in chemically pristine gas, and may not necessarily rule out all pristine gas formation scenarios. 

One thing worth highlighting is an implicit tension in the scenario where Lyman-Werner radiation from a nearby star-forming region is required so as to allow only atomic cooling to prevent fragmentation and thus allow the formation of a supermassive star or a DCBH from pristine gas. Recently it has been argued that at least 43\% of LRDs currently have a blue emitting `blob' close enough to the LRD to provide this radiation \citep{Baggen2026-blobs}.  At the same time, our results show that most LRD hosts are metal-enriched. For the blue blobs near the LRDs to have prompted the collapse of the object from pristine gas, the metal enrichment would have to happen after the blue blob starts emitting, with the enrichment timescale being significantly shorter than the lifetime of the star-forming blob, so that the blob is still observable. 

\subsection{Comparison with high redshift star forming galaxies}
\label{sec:discussion-sf-or-agn}

In Fig.~\ref{fig_discussion: redshift evolution} we show the direct $T_\mathrm{e}$ oxygen
abundances and electron temperatures derived in this work as a function of
redshift, overlaid on a compilation of direct $T_\mathrm{e}$ measurements for high-$z$
star-forming galaxies selected based on the detection of the \oiiiaur line
\citep{Laseter2024, Morishita2024, Sanders2024, Scholte2025, Pollock2026, Sanders2026}. To guide the comparison, we also plot the metallicity evolution
predicted by the mass-metallicity relation (MZR) of \citet{Sanders2021} for a
range of stellar-mass bins ($\log M_*= 8-10$) down to $z\sim0$. 

The oxygen abundances and electron temperatures of our sample overlap with those
of the general star forming population at the same redshifts: the bulk of the LRDs fall
within the locus traced by metal-poor star-forming galaxies, with temperatures
in the $T_\mathrm{e} \approx 15000{-}30000$\,K range that are characteristic of these
systems. In this sense, the narrow components of LRDs are
fully compatible with metal-poor star formation. We note, however, a systematic
tendency: at a given redshift the LRDs preferentially occupy the
lower metallicity/higher temperature edge of the observed distribution, rather than the average. 

Interestingly, the deviation of the LRDs
from the typical abundance of star-forming galaxies is smallest at $z\sim7$ and
grows steadily as redshift decreases. This deviation becomes most significant for the local LRDs discovered in \citet{Lin2026-DESI,Lin2025}, which occupy the same part of parameter space as the high-$z$ sample and do not follow the trend of the MZR relation at $z < 2$, visualized in Fig.~\ref{fig_discussion: redshift evolution}. This has major implications for the evolution of LRDs across cosmic time; if LRD hosts continued to form stars then there would have to be some level of evolution of their metallicity with time. Consequently, either star-formation is completely quenched, or these objects are observed as LRDs at a certain stage of their lifetime. We note that this lack of evolution of the metallicity of LRDs across cosmic time persists even if one adopts $n_\mathrm{e} \sim 10^5\,\,\mathrm{cm^{-3}}$ for our sample.

The agreement between our direct-$T_\mathrm{e}$ abundances and the strong-line estimates
reported for individual LRDs is also informative. Strong-line
calibrations are directly tied to star-formation physics, so their broad
consistency with the direct method indicates that the narrow-line component of
LRDs is, to first order, governed by metal-poor star formation. At the same
time, the highest temperatures we infer ($T_\mathrm{e}\gtrsim3\times10^{4}$\,K) are
difficult to produce by star formation alone and point to an additional, harder
ionising source. This is most evident for rubies-egs61\_4233\_55604 and
rubies-egs63\_4233\_49140 ($T_\mathrm{e}\gtrsim30000$\,K), which stand out as outliers relative
to both the rest of the LRD sample and the $z\sim7$ comparison galaxies, and may be explained by a contribution from AGN ionisation. A similar argument can be invoked for the LRDs in the sample for which the $T_\mathrm{e}$ measurement is unconstrained, due to a low \oiiidoublet/\oiiiaur ratio. Taken together, the
data hint at a composite scenario in which perhaps both metal-poor star formation and an
AGN contribute to the observed line emission on an object-to-object basis. If this is the case, the agreement between the direct and strong-line metallicities further implies that the host galaxy and the AGN narrow-line region of LRDs share similar enrichment properties, i.e.\ both are likely to be similarly metal-poor.


\subsection{Do LRDs deviate from the FMR?}
\begin{figure}[h]
    \centering
    \includegraphics[width=0.45\textwidth]{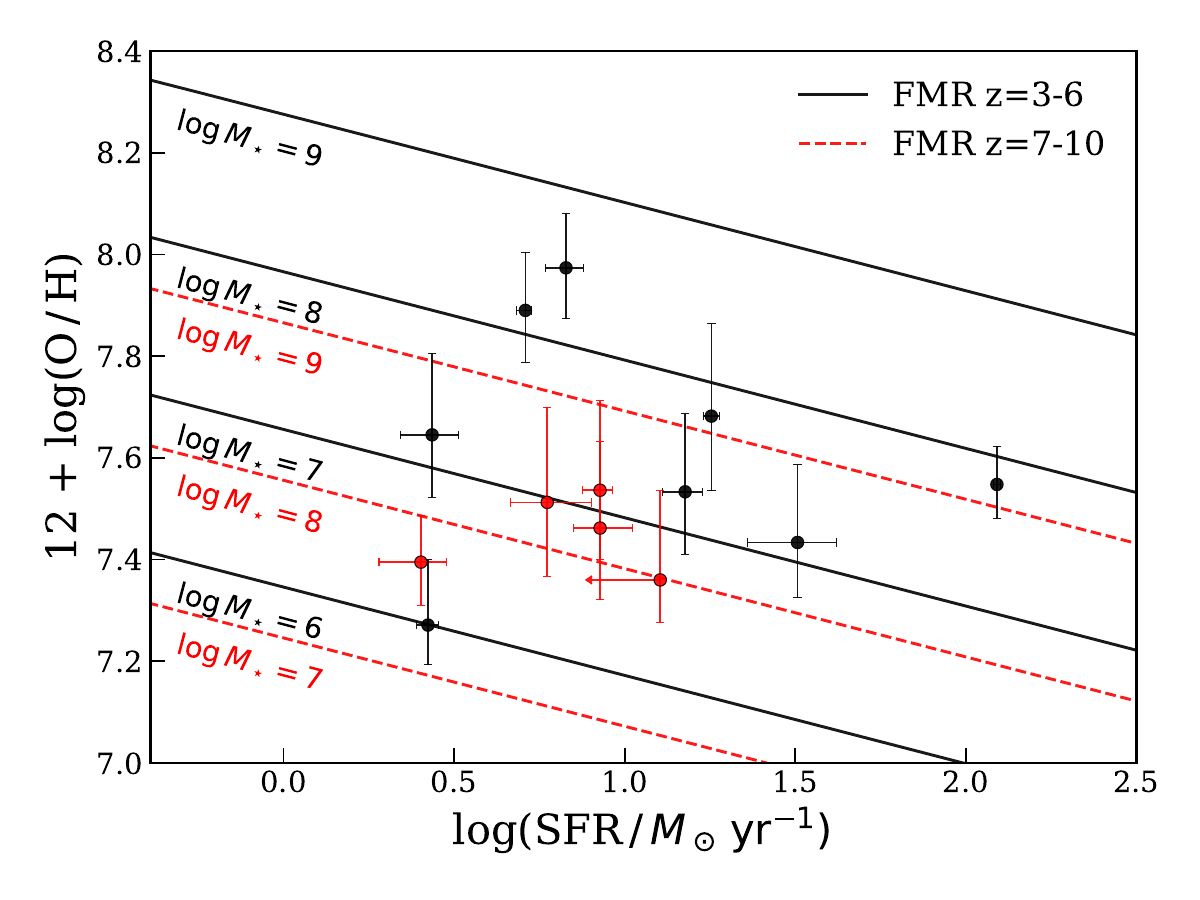}
       \caption{Metallicities vs. star-formation rate for LRDs with well-constrained metallicities (from the direct-\(T_\mathrm{e}\) method). We overplot the FMR from \citet{Curti2024} at $z = 3{-}6$ (black solid lines) and $z = 7{-}10$ (red dashed lines), for different stellar masses. The FMR-inferred stellar masses for LRD hosts are quite low, \(\log{M_*/\mathrm{M}_\odot}\!\sim\!8\).}
          \label{fig_discussion: FMR}
\end{figure}

Figure~\ref{fig_discussion: FMR} sees our LRD sample placed on the
log\,SFR(\ha)-12$+$log(O/H) plane, overlaid with the
\citet{Curti2024} parameterisation of the Fundamental Metallicity Relation (FMR) at \(z\!=\!3{-}6\) and at \(z\!=\!7{-}10\), 
evaluated at stellar masses \(\log{M_*/\mathrm{M}_\odot}\!=\!6{-}10\). All LRD hosts with a direct $T_\mathrm{e}$ measurement from Table~\ref{tab:metallicity_te} and a star formation rate (SFR) estimate using the dust-corrected narrow \ha flux (where applicable) are shown. SFRs are computed using the metal-poor conversion factor from \citet{Reddy2022, Shapley2023}.

Our $z\!<\!6$ sources lie mostly around the \(z\!=\!3{-}6\) FMR at stellar masses of \(\log{M_*/\mathrm{M}_\odot}\!=\!7{-}8.5\), with 4/7 LRD stellar masses consistent with \(\log{M_*/\mathrm{M}_\odot}\!\sim\!7.5\), as well as a lower mass outlier down at \(\log{M_*/\mathrm{M}_\odot}\!\sim\!6\).
The higher-redshift (\(z\!\geq\!6\)) subset is slightly lower in metallicity at fixed SFR, by \(<\!0.5\)\,dex and corresponds to \(\log{M_*/\mathrm{M}_\odot}\!=\!8{-}9\) in the \(z\!=\!7{-}10\) relation. 
Given that our median redshift in the lower redshift set is 4.6, but only 6.7 for the higher redshift set, they may be a little closer in inferred stellar mass than the above suggests at face value. Regardless, these stellar masses are generally lower than the stellar masses derived in the literature for LRDs, which have mostly relied on SED fitting to what is likely nebular emission from the LRD itself \citep{Sneppen2026b}, and can therefore be treated effectively as upper limits. 
Estimates vary significantly (\(\log{M_*/\mathrm{M}_\odot}\!=\!7{-}11\)), depending on the adopted model \citep[e.g.][]{Leung2024, Kocevski2025}; although see \citet{Juodzbalis2026-Qso1} for a dynamical mass upper limit of \(\log{M_*/\mathrm{M}_\odot}\!<\!7.3\) for the host galaxy of abell2744-spurs02\_9214\_41. 
Assuming that LRD hosts follow the FMR at their respective redshift, their implied stellar masses (\(\sim\!10^8\)\,\msun) give very high black hole-to-stellar mass ratios: \(M_\mathrm{BH}/M_*\!\gtrsim\!0.01\), assuming SMBH masses in the range \(\log{M_\mathrm{BH}}\!=\!6{-}7\),  with ratios larger than unity not ruled out. For example, adopting 
\(\log{M_\mathrm{BH}/\mathrm{M}_\odot}\!\sim\!6\) for the \(\log{M_*/\mathrm{M}_\odot}\!\sim\!6\) outlier, jades-gdn\_1181\_73488 \citep{Rusakov2025}, gives \(M_\mathrm{{BH}/M_*}\!\simeq\!1\). This result suggests either that LRDs host SMBHs that, despite their lower masses inferred after correcting for electron-scattering broadening, may still be overmassive with respect to their hosts, or that they deviate from the FMR at their respective redshift. To be consistent with the FMR at a given redshift, would require metallicities 2--3 times what we observe.

Observations at high redshift show a systematic offset from the FMR at $z\!\sim\!0$; by $-\,0.2$\,dex at $z\!\sim \!3$ down to $-\,0.5$\,dex at $z\!\sim\!10$, with significant scatter \citep{Pollock2026}. This offset in log(O/H) suggests that for a given metallicity and star formation rate, high-$z$ galaxies are more massive than predicted by the low-$z$ FMR by $>\!1$\,dex. That LRDs could be offset from their local FMR has recently been discussed by \citet{Isobe2026} for two LRDs, while \citet{Ivey2026} discuss a possible deviation from the mass-metallicity relation for The Cliff ($z = 3.55$, \citealt{deGraaff2025}).
As we showed in section~\ref{sec:discussion-sf-or-agn}, LRDs are less enriched than other star-forming galaxies at their respective redshifts, hinting that they may indeed systematically deviate from the high-$z$ FMRs, if their stellar masses are not found to be systematically less than inferred from SED fitting analyses. 
A definitive test would require independent dynamical mass and black hole mass estimates for individual sources.

\section{Summary}
\label{sec: conclusion}
We have presented the first systematic study of the gas-phase metallicities of Little Red Dots based on medium and high resolution JWST/NIRSpec spectroscopy, providing a catalogue of oxygen abundances for a sample of 24 LRDs at $z \approx 2.3-7$, while constraining the electron temperature in 13 objects. We find that: 
\begin{itemize}
    \item Using the direct $T_\mathrm{e}$ method applied to the narrow components of observed lines, we recover a sample averaged oxygen abundance of $Z_{T_\mathrm{e}} = 0.08_{-0.03}^{+0.11}\,\mathrm{Z_{\odot}}$ with an average electron temperature of $T_\mathrm{e} = 23000_{-7000}^{+17000}$. LRD hosts are therefore metal-poor, but not chemically pristine. 
    \item The \rhat strong line calibration yields a fully consistent sample average of $Z_{\hat{\mathrm{R}}} = 0.07_{-0.04}^{+0.07}\,\mathrm{Z_{\odot}}$, with the 1$\sigma$ $T_\mathrm{e}$-to-\rhat metallicity scatter being just 4\% in 12+log(O/H) space. The \rhat calibration can thus be used to safely constrain the oxygen abundances of LRD hosts. Combining the direct $T_\mathrm{e}$ and N2-based upper limits on metallicity yields a joint average of $Z_{\mathrm{joint}} = 0.08_{-0.04}^{+0.09}\,\mathrm{Z_{\odot}}$.
    \item We find two very metal poor LRDs in the sample, abell2744-spurs02\_9214\_41 (A2744-QSO1 $z\sim7.04$), with  $Z\,<\,1.3$\%\,$\mathrm{Z_{\odot}}$, and jades-gds-w09\_1212\_2993 ($z\!\sim\!4.88$), with \(Z\!=\!0.3\)\%\,$\mathrm{Z_{\odot}}$, constrained using the \rhat calibration. These are the only two objects in the sample that approach the abundances theoretically required for DCBH-based formation. 
    \item Comparing grating and PRISM derived abundances, we find large object-to-object scatter of up to $\sim\!0.8$\,dex at $1\sigma$. PRISM spectra at the usual redshift range of LRDs, lack the resolution to sufficiently decompose the \hb+\oiiidoublet, and \hg+\oiiiaur complexes, and PRISM-based direct $T_\mathrm{e}$ metallicities of samples of LRDs should be treated with caution.
    \item The metallicities and temperatures of LRDs are within the range of the general metal-poor star forming population at the same redshifts, but preferentially occupy the lower-metallicity/higher-temperature edge of the distribution. \item LRD metal abundances remain remarkably constant across cosmic time, suggesting that low metallicity is a defining characteristic of LRDs. 
    \item Placing LRDs on the FMR implies low stellar masses, \(\log{M_*/\mathrm{M_\odot}}\!\sim\!7{-}9\). The average black hole masses quoted for LRDs in the literature either imply that they host SMBHs that are overmassive for their hosts, or that they deviate from the FMR at their respective redshifts.   
\end{itemize}

Taken together, our results indicate that the narrow line emission in LRDs is largely consistent with a metal-poor star-forming host, though a subset shows signatures of an additional harder ionising source. In a scenario where the narrow emission is a composite of a host galaxy and an AGN, our results imply that the enrichment of the two components is similar. Probing the chemical enrichment of the cocoon around the central engine is an important future endeavour to discriminate for certain that the LRD and its host share the same metal abundance. This might be done with oxygen or iron permitted lines with a suitable modelling of the Ly\(\beta\) radiative transfer or with observations of semi-forbidden lines with high critical density, possibly in the rest-frame UV. Finally, to truly paint a coherent picture of the nature of LRDs and establish their relation to the FMR and their black hole to host galaxy mass ratio, will require independent dynamical and black-hole mass estimates for several individual sources \citep[e.g.][]{Juodzbalis2026-Qso1}.
\section*{Data availability}
This paper makes use of public JWST data downloaded from the DAWN JWST Archive (DJA)\footnote{DOI: 10.5281/zenodo.8319596}.
\begin{acknowledgements} 
The Cosmic Dawn Center (DAWN) is funded by the Danish National Research Foundation under grant DNRF140. The data products presented herein were retrieved from the DAWN JWST\ Archive (DJA). DJA is an initiative of the Cosmic Dawn Center. DW, GPN and AS are co-funded by the European Union (ERC, HEAVYMETAL, 101071865). Views and opinions expressed are, however, those of the authors only and do not necessarily reflect those of the European Union or the European Research Council. Neither the European Union nor the granting authority can be held responsible for them. KEH acknowledges funding from the Swiss State Secretariat for Education, Research and Innovation (SERI). This work is based in part on observations made with the NASA/ESA/CSA James Webb Space Telescope. The data were obtained from the Mikulski Archive for Space Telescopes (MAST) at the Space Telescope Science Institute, which is operated by the Association of Universities for Research in Astronomy, Inc., under NASA contract NAS 5-03127 for JWST. These observations are associated with programmes \#1180, \#1181, \#1286, \#1212, \#9223, \#3567, \#4233, \#1345, \#4106, \#4287 and \#9214.
 
\end{acknowledgements}
\bibliographystyle{mnras}
\bibliography{aanda} 

\appendix 
\onecolumn
\section{Flux measurements}

\begin{table}[!ht]
\centering
\renewcommand{\arraystretch}{1.25}
\setlength{\tabcolsep}{5pt}
\caption{Narrow line fluxes for the LRD sample}
\label{tab:fluxes}
\resizebox{0.98\textwidth}{!}{%
\begin{tabular}{@{}clcccccccc@{}}
\hline
 ID & Object & \(z\) & F(\oiii D) & F(\oiii) &  F(\oii) & F(\neiii) & F(\nii) & F(\ha ) & F(\hb ) \\ \hline
1 & abell2744-spurs02\_9214\_41 & $7.037$ & $<5$ & $<2$ & $<3$ & $<3$ & $<4$ & $<72$ & $3^{+1}_{-1}$  \\
2 & rubies-egs61\_4233\_55604 & $6.983$ & $147^{+2}_{-2}$ & $9^{+1}_{-1}$ & $<10$ & $14^{+2}_{-2}$ & $<3$ & $<62$ & $14^{+3}_{-5}$  \\
3 & rubies-uds42\_4233\_807469 & $6.775$ & $72^{+2}_{-2}$ & $<7$ & $<8$ & $4^{+1}_{-1}$ & $<3$ & $<30$ & $9^{+2}_{-2}$  \\
4 & jades-gdn2\_1181\_954 & $6.760$ & $150^{+3}_{-3}$ & $5^{+1}_{-1}$ & $<7$ & $7^{+1}_{-1}$ & $<3$ & $36^{+3}_{-4}$ & $14^{+2}_{-2}$  \\
5 & egs-nelsonx\_4106\_47962 & $6.728$ & $150^{+2}_{-2}$ & $4^{+1}_{-1}$ & $<13$ & $7^{+2}_{-1}$ & $<4$ & $37^{+9}_{-6}$ & $21^{+2}_{-2}$  \\
6 & rubies-egs63\_4233\_49140 & $6.685$ & $143^{+2}_{-2}$ & $7^{+2}_{-2}$ & $6^{+2}_{-2}$ & $12^{+2}_{-2}$ & $<2$ & $26^{+9}_{-6}$ & $12^{+3}_{-3}$  \\
7 & glimpse-obs01b\_9223\_5536 & $6.223$ & $11.0^{+0.4}_{-0.4}$ & $<1$ & --- & --- & $<2$ & $<6$ & $1.1^{+0.4}_{-0.4}$  \\
8 & glimpse-obs01\_9223\_12248 & $6.108$ & $65^{+1}_{-1}$ & $1.8^{+0.3}_{-0.3}$ & --- & --- & $<1$ & $14^{+3}_{-3}$ & $9.6^{+0.4}_{-0.4}$  \\
9 & jades-gdn\_1181\_38147 & $5.869$ & $259^{+3}_{-3}$ & $7^{+2}_{-2}$ & $16^{+3}_{-3}$ & $9^{+2}_{-2}$ & $<9$ & $89^{+11}_{-12}$ & $28^{+2}_{-2}$  \\
10 & rubies-uds3\_4233\_47509 & $5.672$ & $199^{+3}_{-3}$ & $<9$ & --- & --- & $<3$ & $73^{+9}_{-20}$ & $20^{+3}_{-3}$  \\
11 & ceers\_1345\_746 & $5.622$ & $22^{+1}_{-1}$ & $<9$ & $<5$ & $<5$ & $<3$ & $8^{+3}_{-2}$ & $<5$  \\
12 & jades-gds05\_1286\_204851 & $5.482$ & $138^{+2}_{-2}$ & $<7$ & $7^{+2}_{-2}$ & $7^{+2}_{-1}$ & $<1$ & $24^{+5}_{-5}$ & $12^{+2}_{-2}$  \\
13 & ceers-egs\_1345\_2782 & $5.239$ & $96^{+2}_{-2}$ & $<6$ & $<7$ & $5^{+1}_{-1}$ & $<2$ & $<45$ & $10^{+3}_{-3}$  \\
14 & jades-gds06\_1286\_159717 & $5.077$ & $552^{+6}_{-6}$ & $15^{+4}_{-4}$ & $56^{+9}_{-9}$ & $32^{+6}_{-6}$ & $<7$ & $149^{+8}_{-8}$ & $48^{+5}_{-5}$  \\
15 & jades-gdn\_1181\_68797 & $5.039$ & $370^{+22}_{-26}$ & $17^{+3}_{-3}$ & $59^{+6}_{-6}$ & $30^{+3}_{-3}$ & $<30$ & $272^{+81}_{-77}$ & $37^{+7}_{-10}$  \\
16 & valentino-egs\_3567\_42232 & $4.952$ & $25^{+1}_{-1}$ & $<4$ & $<6$ & $3^{+1}_{-1}$ & $<2$ & $26^{+6}_{-7}$ & $4^{+1}_{-1}$  \\
17 & jades-gdn\_1181\_39353 & $4.849$ & $58^{+2}_{-2}$ & $<4$ & $<7$ & $<8$ & $<2$ & $32^{+4}_{-7}$ & $9^{+2}_{-2}$  \\
18 & jades-gds-w09\_1212\_2993 \tablefootmark{1}& $4.824$ & $9^{+2}_{-2}$ & $<17$$\,$& $<10$& $7^{+2}_{-2}$ & $<18$ & $91^{+6}_{-6}$ & $19^{+2}_{-2} $  \\
19 & jades-gds02\_1286\_38562 & $4.821$ & $196^{+2}_{-2}$ & $3^{+1}_{-1}$ & $10^{+1}_{-1}$ & $10^{+1}_{-1}$ & $<2$ & $48^{+2}_{-3}$ & $18^{+1}_{-1}$  \\
20 & ceers-egs\_1345\_1244 & $4.477$ & $494^{+19}_{-18}$ & $7^{+1}_{-1}$ & $23^{+3}_{-3}$ & $21^{+2}_{-2}$ & $5^{+2}_{-2}$ & $75^{+9}_{-10}$ & $44^{+6}_{-5}$  \\
21 & jades-gdn\_1181\_73488 & $4.133$ & $152^{+1}_{-1}$ & $6^{+1}_{-1}$ & $<7$ & $7^{+1}_{-1}$ & $<3$ & $36^{+3}_{-3}$ & $22^{+1}_{-1}$  \\
22 & jades-gds-wide\_1180\_13329 & $3.938$ & $190^{+3}_{-3}$ & $<9$ & $20^{+5}_{-5}$ & $11^{+3}_{-3}$ & $<7$ & $63^{+5}_{-7}$ & $16^{+3}_{-2}$  \\
23 & jades-gdn\_1181\_53501 & $3.429$ & $282^{+4}_{-4}$ & $11^{+3}_{-2}$ & $<22$ & $<24$ & $<3$ & $57^{+11}_{-11}$ & $18^{+4}_{-4}$  \\
24 & jades-gdn2\_1181\_28074 & $2.260$ & $15449^{+194}_{-214}$ & $352^{+42}_{-45}$ & $2747^{+61}_{-61}$ & $2014^{+39}_{-38}$ & $87^{+22}_{-23}$ & $7080^{+174}_{-175}$ & $2392^{+111}_{-107}$ \\ \hline
\end{tabular}
}\tablefoot{All fluxes are given in units of $10^{-19}\mathrm{\,ergs\,s^{-1}\,cm^{-2}}$, and have been corrected for dust where applicable. F(\oiii D) denotes the flux in the \oiiidoublet doublet,  F(\oiii) corresponds to the flux in the \oiiiaur line, while F(\oii), F(\neiii), and F(\nii) correspond to the fluxes in the \oiidoublet doublet, \neoniii, and \niired lines respectively.\\
\tablefoottext{1}{The \oiiidoublet, \oiidoublet, \neoniii and \hb fluxes quoted for this object are estimated using PRISM spectra.}}
\end{table}

\end{document}